\documentclass[twocolumn,showpacs,preprintnumbers,amsmath,amssymb,superscriptaddress]{revtex4-1}

\usepackage{graphicx}
\usepackage{dcolumn}
\usepackage{bm}
\usepackage{graphics}
\usepackage{subfigure}
\usepackage{graphicx}
\usepackage{epsfig}
\usepackage{epstopdf}
\usepackage{color}

\begin{document}

\title{Pairing effects in the nondegenerate limit of the two-dimensional Fermi gas}

\author{Marcus Barth}
\email{marcus.barth@ph.tum.de}
\affiliation{Physik\,Department,\,Technische\,Universit{\"a}t\,M{\"u}nchen,
James-Franck-Strasse,\,85748 Garching,\,Germany}
\author{Johannes Hofmann}
\email{hofmann@umd.edu}
\affiliation{
Condensed Matter Theory Center, Department of Physics, University of Maryland, College Park, Maryland 20742-4111 USA
}
\date{\today}

\begin{abstract}
The spectral function of a spin-balanced two-dimensional Fermi gas with short-range interactions is calculated by means of a quantum cluster expansion. Good qualitative agreement is found with a recent experiment by Feld \textit{et al.} [Nature (London) \textbf{480}, 75 (2011)]. The effects of pairing are clearly visible in the density of states, which displays a suppression of spectral weight due to the formation of a two-body bound state. In addition, the momentum distribution and the radio-frequency spectrum are derived, which are in excellent agreement with exact universal results. It is demonstrated that in the limit of high temperature, the quasiparticle excitations are well defined, allowing for a kinetic description of the gas.
\end{abstract}

\maketitle

\section{Introduction}

Feshbach resonances in ultracold atoms provide us with the possibility of tuning the strength of the interparticle interaction at will, allowing us to probe vastly different types of physics. For a Fermi gas at low temperature, this ranges from a BCS-type superfluid at small attractive interaction to a Bose-Einstein condensate (BEC) of tightly bound dimers as the interaction strength is increased. This BEC-BCS crossover has been the subject of intense research over the past decade~\cite{zwerger12}. An interesting question is whether pairing affects the properties of a Fermi gas above the superfluid transition temperature as well. In contrast to standard BCS theory, which predicts pairing and condensation to appear simultaneously, it has been conjectured that pairing occurs at a temperature larger than the superfluid transition temperature, and that the remnant of a pairing gap remains in the normal phase. This regime is known as the pseudogap phase. It is expected that in the pseudogap phase, the 
single-particle excitation 
spectrum assumes a BCS-type dispersion relation $\omega({\bf q}) = \sqrt{(\varepsilon_{\bf q}-\mu)^2+\Delta^2}$, where $\varepsilon_{\bf q} = q^2/2m$, $\mu$ is the chemical potential, and $\Delta$ is a superfluid order parameter, which predicts a ``back-bending'' of the dispersion relation around the Fermi momentum. For a three-dimensional unitary Fermi gas, the single-particle excitation spectrum has been probed using momentum-resolved radio-frequency spectroscopy~\cite{stewart08}, and evidence of pseudogap behavior has been reported~\cite{gaebler10}. Various theoretical works indicate the existence of a pseudogap~\cite{magierski09,chen09}, while some others do no observe this~\cite{haussmann09}. Generally, fluctuations are more relevant in two-dimensional (2D) systems, suggesting that pseudogap effects are more pronounced in 2D. Indeed, Feld \textit{et al.} recently reported the observation of a pairing pseudogap in a two-dimensional Fermi gas~\cite{feld11}. 

Experimentally, quasi two-dimensional Fermi gases are created by trapping the system in a strongly oblate trapping geometry.  For a harmonic trapping potential, the strength of the confinement is set by the ratio of the harmonic oscillator length in the confining direction, $l_z = \sqrt{1/m\omega_z}$, and the 2D scattering length $a_{\rm 2}$, which is related to the 3D scattering length via the transcendental equation $l_z / a_{\rm 3} = f_1 (l_z^2/a_2^2)$. The function $f_1$ is for example given in Sec. V of Ref.~\cite{bloch08}. In this paper, we set $\hbar = k_B = 1$. In the limit of strong confinement in which $l_z$ is much smaller compared to $a_{\rm 2}$, the perpendicular degree of freedom decouples from the dynamics, rendering the system effectively two-dimensional. Note that in contrast to the 3D case, $a_{\rm 2}$ is always positive and there exists a two-body bound state with binding energy $E_b=1/ma_{\rm 2}^2$ for all scattering lengths. In recent years, it has become possible to prepare and probe 
Fermi gases in the strictly two-dimensional regime~\cite{froehlich11,sommer12}, and we restrict our attention to this purely 2D case.

Momentum-resolved rf spectroscopy induces a transition from an initial occupied spin state to an unoccupied state of same momentum, followed by a time-of-flight measurement to extract the momentum distribution of the out-coupled atoms. This transition rate is directly related to the spectral function, which encodes the single-particle excitation spectrum. While the spectral function is fundamental to the description of many-body systems, it is usually a very challenging and complex task to calculate this quantity theoretically. Quite generally, the analysis of strongly interacting Fermi gases is complicated by the lack of a small parameter which could be used in a perturbative expansion, and in many cases, one has to resort to complex numerical calculations to obtain quantitatively reliable results. The experiment~\cite{feld11} has thus far been analyzed using different resummation schemes for the spectral function~\cite{klimin12,pietila12,watanabe13}. In this paper, we apply a quantum 
cluster expansion to the spectral function, which provides a systematic expansion about the nondegenerate or high-temperature limit. This virial expansion has already been successfully applied to extract thermodynamic properties of the two-dimensional gas in a trap~\cite{liu10} as well as the spectral function in the three-dimensional gas~\cite{hu10}. We find that even the leading order provides a qualitative description of the measured data~\cite{feld11}. We discuss the properties of the spectral function in detail, focusing in particular on the density of states and the implications of an incoherent spectral weight found at negative frequencies.

This paper is structured as follows: In Sec.~\ref{sec:spectral}, we start by discussing the phenomenology of the spectral function and its characteristic behavior throughout the BCS-BEC crossover. Section~\ref{sec:virial} introduces the virial expansion. It is established that the virial expansion should be quantitatively reliable up to temperatures as low as the Fermi temperature, and we outline how the spectral function is calculated within the virial expansion. Section~\ref{sec:results} presents the results of this calculation. While the onset of a gap is clearly visible in the density of states, we argue that the observed backbending of the lower branch is not a feature of the spectral function, but a consequence of the asymmetric structure of the bound state branch and the thermal occupation of states. The density of states is presented in Sec.~\ref{sec:dos}. Section~\ref{sec:rfandmomdis} extracts the momentum distsribution and the rf transition rate from the spectral functions. Our 
results 
reproduce known universal results valid in the high-momentum and high-frequency limits, respectively, which are linked to the incoherent weight of the spectral function at large and negative frequency. Furthermore,  in Sec.~\ref{sec:qp}, we compute the quasiparticle properties. The paper is concluded by a summary in Sec.~\ref{sec:conclusion}.

\section{The spectral function}\label{sec:spectral}

We begin by discussing the properties of the spectral function, which contains information about the single-particle spectrum. The spectral function is defined as the imaginary part of the retarded single-particle Green's function:
\begin{align}
\label{eq:spectralfunction-definition}
A(\omega, {\bf q}) &= - 2 \, {\rm Im} \, G(\omega, {\bf q}) .
\end{align}
It describes the probability density of creating either a particle or a hole excitation with momentum ${\bf q}$ and energy $\omega$. For a noninteracting gas with dispersion $\omega({\bf q})$, the spectral function is a $\delta$-function centered at $\omega({\bf q})$. In the presence of interactions, the peak acquires a finite width, which is proportional to the inverse lifetime of the excitation. As an example, consider the Fermi gas at low temperature and large scattering length, i.e., in the BCS regime. The spectral function takes the form
\begin{align}
A(\omega, {\bf q}) &= 2\pi v_{\bf q}^2 \delta(\omega + \sqrt{(\varepsilon_{\bf q} - \mu)^2 + \Delta^2}) \nonumber \\
&+ 2 \pi u_{\bf q}^2 \delta(\omega - \sqrt{(\varepsilon_{\bf q} - \mu)^2 + \Delta^2}) .
\end{align}
Here, $v_{\bf q}$ and $u_{\bf q}$ are the Bogoliubov parameters, and the chemical potential is positive $\mu=k_F^2/2m>0$, whereby the Fermi momentum $k_F$ is related to the density $n$ via $k_F=\sqrt{2\pi n}$. For any fixed momentum, it is not possible to create an excitation in the energy range between $\pm \Delta$: the single-particle spectrum is gapped. Note that the hole-part of the spectral function, which starts at $\omega = - \sqrt{\mu^2 + \Delta^2}$, bends back at $q = k_F$ towards negative frequency at large momentum. Interactions are expected to renormalize the scale at which the backbending occurs to some $k_0 \neq k_F$.

As the strength of the interaction is increased, there is a crossover from the BCS to the BEC regime, in which the quasiparticles are not Cooper pairs but two-particle bound states. In this limit, the spectral function is
\begin{align}
A(\omega,{\bf q}) &= 2 \pi Z_{\bf q} \, \delta(\omega + \varepsilon_{\bf q} - \mu) \nonumber \\ & + 2\pi (1 - Z_{\bf q})  \, \delta(\omega - \varepsilon_{\bf q} + \mu) ,
\end{align}
where the chemical potential is half the bound-state energy, $\mu = - E_b/2$, and the residue is $Z_{\bf q} = |\varphi(q)|^2 n$, with $|\varphi(q)|^2 = 4 \pi a_2^2/(1+q^2 a_2^2)^2$ being the square of the boundstate wave function in momentum space. Note that there is no backbending at finite momentum in the dispersion relation, but the spectral function still possesses a gap of size $E_b$~\cite{randeriabound89}. 

\begin{figure}
\scalebox{0.7}{\includegraphics{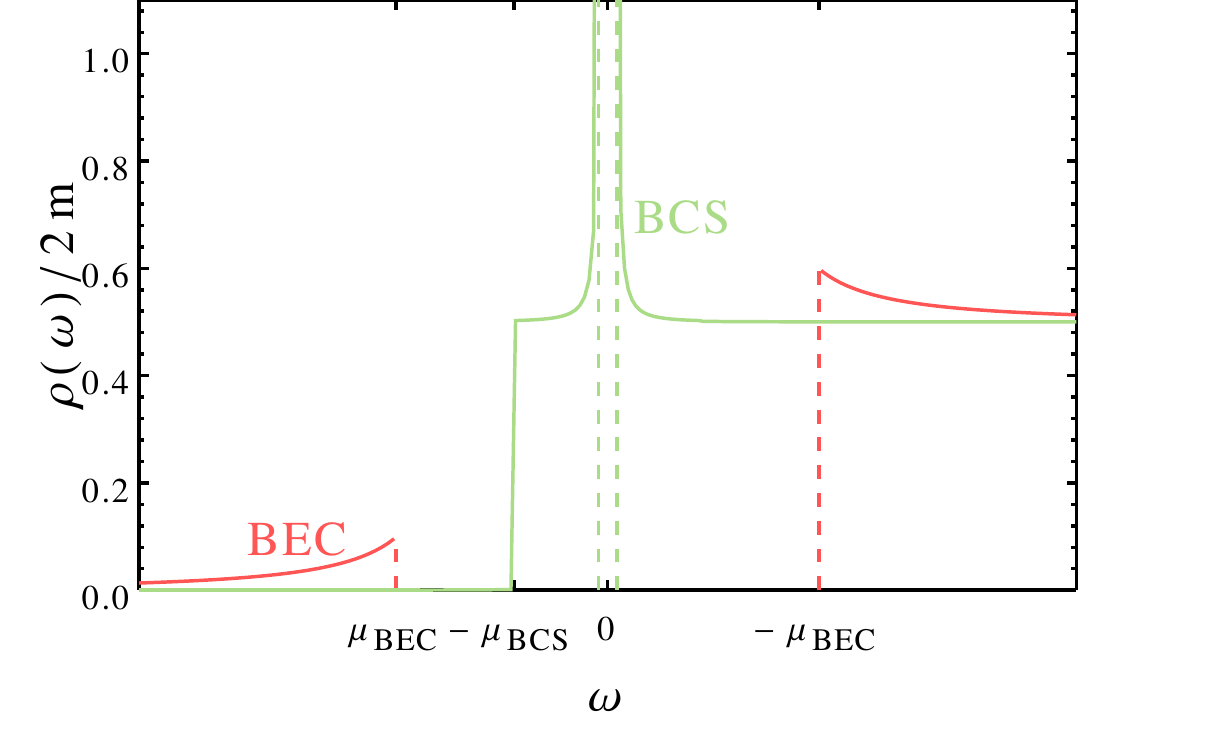}}
\caption{(Color online) Sketch of the density of states at zero temperature in the BCS and the BEC limits.}
\label{fig:dossketch}
\end{figure}

The pairing gap is also manifested in the density of states. The density of states counts the excitations with energy $\omega$, and is obtained by integrating the spectral function over momentum:
\begin{align}
\rho(\omega) &= \int \frac{d^2q}{(2\pi)^2} \, A(\omega, {\bf q}) . \label{eq:dos}
\end{align}
In BCS theory below the superfluid transition temperature, this density of states exhibits a gap of width $2 \Delta$ around the Fermi energy, while in the BEC limit the gap size is given by the binding energy $E_b$ of the two-body bound state. The density of states at zero temperature for both limits is sketched in Fig.~\ref{fig:dossketch}. As the temperature increases beyond the critical temperature, mean-field theory predicts the gap to vanish.

Pairing is possible even above the critical temperature, which can affect the properties of the normal phase. In the high-temperature limit, the quasiparticle excitations are well defined and the fermions are unpaired. Below a certain temperature $T^*$, most fermions are bound in pairs, giving rise to significant deviations from a simple quasiparticle picture. On the BCS side of the crossover, this regime is known as the pseudogap phase. The hallmark of the pseudogap phase is a depletion of spectral weight in the density of states around the Fermi surface at $\omega=\mu$. The pseudogap grows as the temperature is lowered and eventually forms a full gap below $T_c$. Often, a backbending of the dispersion relation akin to that in the BCS model is taken as a phenomenological sign of a pseudogap phase. This, however, has to be treated with caution, since the backbending at large momentum is a generic feature of an interacting Fermi gas, as pointed out by Schneider and Randeria~\cite{schneider10}.
On the BEC side, one also may find a depletion of spectral weight, which in this case is associated with the formation of a two-body bound state and occurs at $\omega = - \mu$. While the pseudogap is considered to be a many-body effect, the depletion on the BEC side can be understood already from a two-body calculation, as will also be shown in Sec. \ref{sec:dos}.

In the BEC limit, the density of fermions $n_f$ and of fermions bound in dimers $n_d$ can be estimated using a thermodynamic argument assuming a noninteracting gas of fermions and dimers~\cite{landau10b,randeria93}. It is given by the so-called Saha formula
\begin{align}
\frac{n_f^2}{n_d} &= \frac{mT}{4\pi} \, e^{- E_b/T} ,
\end{align}
where the total density of particles $n = 2 n_f + 2 n_d$ is kept fixed. We can define a temperature $T^*$ at which there is an equal number of dimers and unpaired fermions in the normal phase of the gas. This temperature is
\begin{align}
\frac{T^*}{T_F} &= \frac{E_b/E_F}{W(E_b/E_F)} , \label{eq:saha}
\end{align}
where $T_F = k_F^2/2m$ denotes the Fermi temperature and $W$ is the Lambert-$W$ function. Below this temperature, fermions are predominantly paired and we expect pronounced pairing effects on the properties of the gas.

Experimentally, the single-particle excitations of cold atomic gases have been measured using momentum-resolved radio-frequency spectroscopy, which is analogous to angle-resolved photoelectron spectroscopy in condensed matter physics~\cite{stewart08,gaebler10,feld11}. The experiment detects the hole excitations, i.e., the rate of transition from occupied to unoccupied states. According to Fermi's golden rule, the transition rate is proportional to
\begin{align}
A_-(\omega, {\bf q}) &= 2 \pi \sum_{n,m} e^{- \beta E_m} \, |\langle n | c_{\bf q} | m \rangle |^2 \, \delta(\omega - E_n + E_m) .
\end{align}
This quantity is known as the {\it occupied spectral function}. Here, the annihilation operator $c_{\bf q}$ destroys a particle with momentum ${\bf q}$.  The full spectral function also includes processes that probe the transition from an unoccupied state to an occupied state if one particle is added to the system. The occupied spectral function is related to the full spectral function by a Fermi-Dirac distribution $f(\omega) = 1/(\exp \beta \omega + 1)$:
\begin{align}
A_-(\omega, {\bf q}) &= f(\omega) A(\omega, {\bf q}) .
\end{align}
In the following, we use a quantum cluster expansion to calculate the spectral function in the nondegenerate limit.

\section{The Virial expansion}\label{sec:virial}
\begin{figure}
\subfigure[\label{subfig:fugacity}]{\scalebox{0.7}{\includegraphics{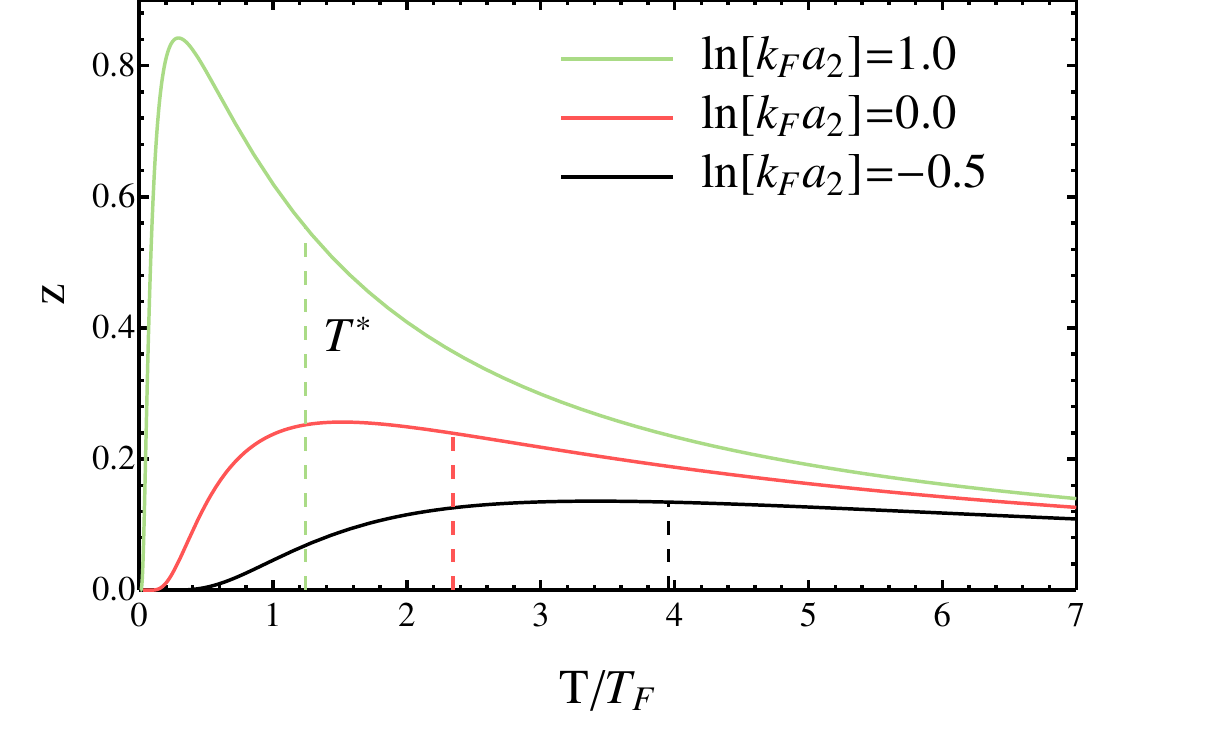}}}
\subfigure[\label{subfig:chempot}]{\scalebox{0.7}{\includegraphics{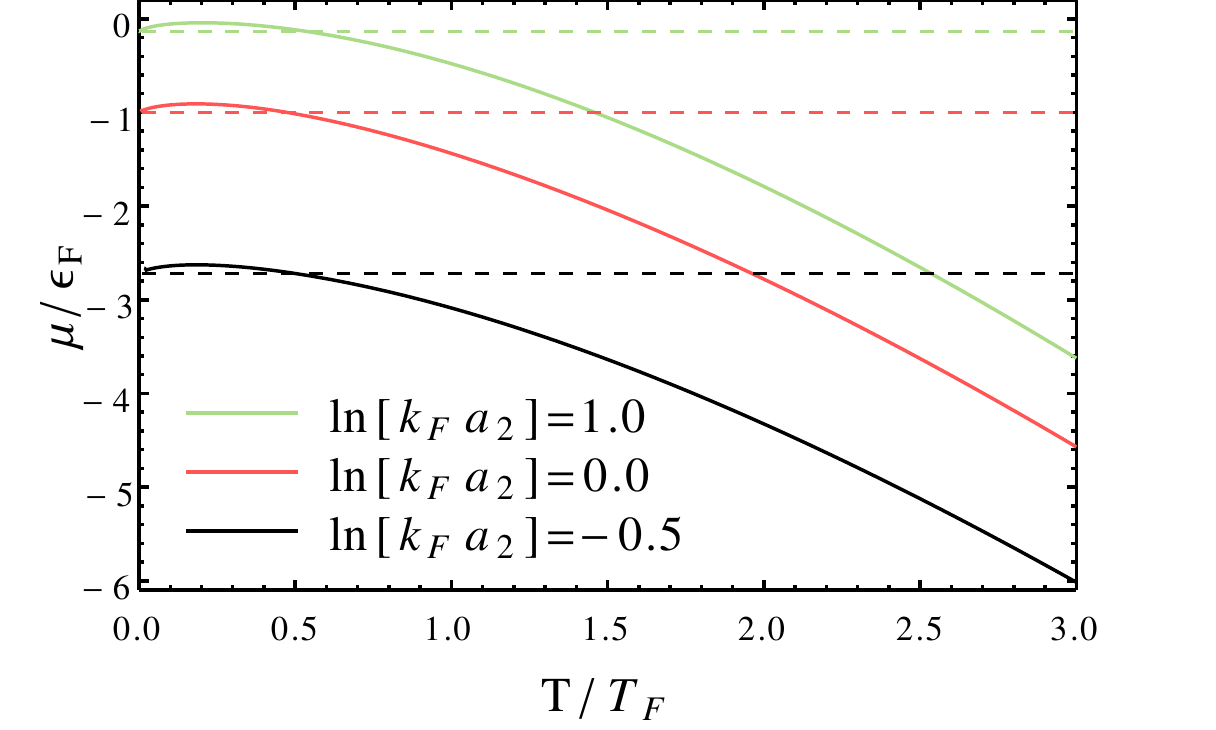}}}
\caption{ (Color online) 
(a) Fugacity as a function of temperature for fixed density. The dashed lines indicate the Saha estimate for $T^*$, Eq.~\eqref{eq:saha}. (b)  Chemical potential as a function of temperature. Dashed lines correspond to half the bound-state energy $-E_b /2 E_F$.}
\label{fig:b2andfug}
\end{figure}

\begin{figure*}
\subfigure[\label{subfig:spectralfunction}]{\scalebox{0.6}{\includegraphics{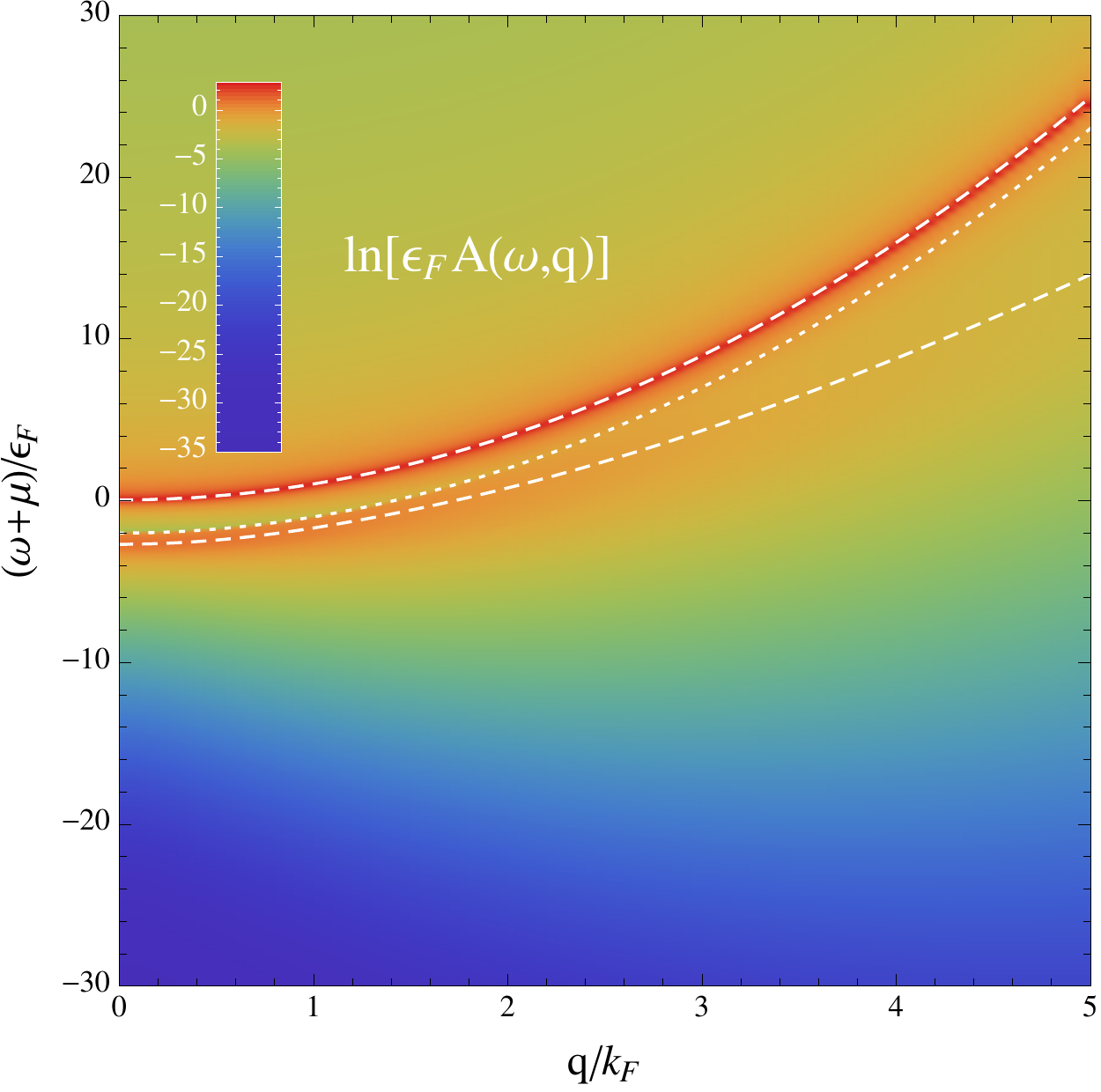}}\hspace*{0.5cm}} \quad
\subfigure[\label{subfig:occspectralfunction}]{\scalebox{0.6}{\includegraphics{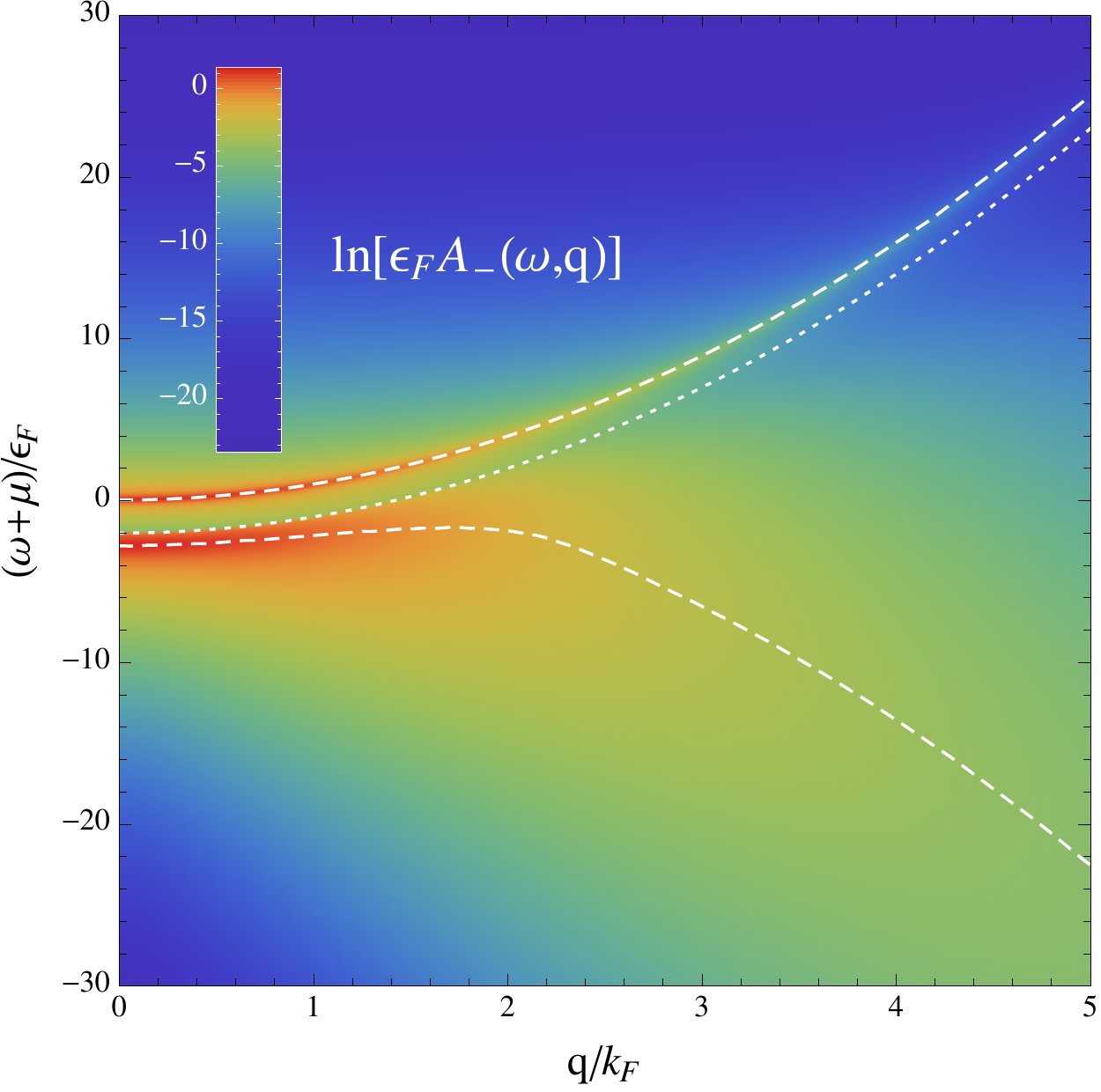}}\hspace*{0.5cm}}
\caption{(Color online) (a) Spectral function at $T=T_F$ and $\ln k_F a_{\rm 2} = 0.0$. (b) Occupied spectral function for the same parameters. The white dashed lines in both figures mark the maxima of the quasiparticle and the lower branch. The white dotted line corresponds to the threshold dispersion $\omega_{\rm th} ({\bf q}) + \mu = - E_b + \varepsilon_{\bf q}$.}
\label{fig:spectralfunction}
\end{figure*}
The virial expansion provides a systematic method for analyzing a Fermi gas at high temperature. The virial expansion is applied to a nondegenerate gas for which the thermal energy $E_T = T$ outweighs its kinetic energy $E_K=\pi n/m$: $E_T \gg E_K$. Equivalently, this corresponds to the limit in which the thermal deBroglie wavelength $\lambda_T = \sqrt{2 \pi/m T}$ is small compared to the interparticle spacing $n^{-1/2}$: $\lambda_T \ll n^{-1/2}$. In this limit, the grand canonical partition function
\begin{align}
\label{eq:paritionfct}
 \mathcal{Z} = \text{tr} e^{-\beta (H - \mu N)} = \sum_{N=1}^{\infty} z^N \text{tr}_N e^{-\beta H}
\end{align}
can be expanded in terms of the fugacity $z=e^{\beta \mu} \ll 1$. The traces $\text{tr}_N$ on the right-hand side of Eq.~\eqref{eq:paritionfct} are restricted to the $N$-particle Fock spaces. Thus, the coefficients of the expansion are determined by clusters that involve one-, two-, and three-body processes, and so on. In this sense, the virial expansion bridges the gap between known few-particle results and the behavior of a complicated many-body system. In particular, the expansion is valid even in a strongly interacting regime. The number density can be obtained directly from Eq.~\eqref{eq:paritionfct}:
\begin{align}
\label{eq:density-grandcanonical}
n &= \frac{2}{\lambda_T^2} \left(b_1 z + 2 b_2 z^2 + \cdots\right) ,
\end{align}
where $b_1$ and $b_2$ are known as the virial coefficients. The prefactor of $2$ counts the two spin species.

For a noninteracting gas, a direct calculation of the virial coefficients gives $b_n^{(0)} = (-1)^{n-1}/n^2$. Interactions enter only in second and higher orders. The correction to the second order is given by the well-known Beth-Uhlenbeck term~\cite{beth37}
\begin{align}
\label{eq:beth-uhl}
\Delta b_2 = b_2 - b_2^{(0)} = \frac{1}{\pi} \int_0^\infty dk \, \frac{\partial \delta(k)}{\partial k} \, e^{- \beta k^2/m} + e^{\beta E_b} ,
\end{align}
where the scattering phase shift is $\cot \delta(k) = (2/\pi) \ln a_{\rm 2} k$. The interaction correction~\eqref{eq:beth-uhl} consists of a bound state contribution and a contribution due to scattering states. The attractive interaction between the particles increases the virial coefficients compared to the noninteracting case and thus tends to increase the density at a given chemical potential. 
In Fig.~\ref{subfig:fugacity}, we show the fugacity as a function of $T/T_F$ as determined from Eq.~\eqref{eq:density-grandcanonical}. The curves for different coupling strengths have a maximum and then tend to zero with decreasing temperature. For comparison, we also include the Saha estimate for $T^*$ as a dashed line in Fig.~\ref{subfig:fugacity}. The virial expansion appears to be valid even in a temperature range below $T^*$. This suggests that the leading-order term is sufficient to quantify pairing effects on the Fermi gas.

In Fig.~\ref{subfig:chempot}, we show the chemical potential for the same coupling strengths as in Fig.~\ref{subfig:fugacity}. For all three of them, the chemical potential exceeds the dimer chemical potential $-E_b/2$ at roughly $T/T_F=0.5$. Since the second-order calculation includes only two-body effects, it should be energetically cheaper for the particles to just form a dimer bound state. This inconsistency provides a clear lower bound $T/T_F=0.5$ on the extrapolation of the second order results to low temperatures.
We regard the virial expansion to be valid down to a temperature of $T_F$, consistent with the findings of other applications of the virial expansion to Fermi gases~\cite{liu10,hu10}. We argue that the results of our virial expansion qualitatively describe the experiment of Ref.~\cite{feld11}, which was carried out in a temperature range $T/T_F = 0.27\mbox{--}0.65$ for scattering lengths in the range between $\ln k_F a_{\rm 2} = -2$ and~$1$. For these interaction strengths, the chemical potential is negative and there exists no Fermi surface~\cite{bertaina11}. The single-particle spectrum should be dominated by the dimer pairing, which is precisely what is captured by the virial expansion.

The starting point of our calculation is the virial expansion of the self-energy, i.e., the one-particle irreducible contribution to the single-particle Green's function. It is related to the Green's function by a Dyson equation~\footnote{In this paper, we suppress spin indices because we are considering a balanced gas. Notice that our Green's functions are not summed over the two indices.},
\begin{align}
\label{eq:GF-dyson}
G(\omega, {\bf q}) &= \frac{1}{\omega + \mu - \varepsilon_{\bf q} - \Sigma(\omega, {\bf q})}.
\end{align}
To linear order in $z$, the self-energy is given by a Boltzmann-weighted integral of the $T$-matrix element (similar to the bosonic case~\cite{nishida13}):
\begin{align}
\label{eq:self-energy-expansion}
 \Sigma^{(1)}(i \omega_n, {\bf q})   & = z  \int \frac{d^2k}{(2\pi)^2} e^{- \beta \varepsilon_{\bf P}} \, T_2(i \omega_n + \mu + \varepsilon_{\bf k}, {\bf k + q}),
\end{align}
where $\omega_n = (2n+1) \pi T$ are fermionic Matsubara frequencies. Equation~\eqref{eq:self-energy-expansion} describes the self-energy correction due to scattering with a single thermally excited particle-hole pair. In the Appendix, we provide a short derivation of this result. To obtain the retarded self-energy, we analytically continue Eq. \eqref{eq:self-energy-expansion} to real frequencies $i \omega_n \to \omega + i0$, replacing the $T$-matrix element in Matsubara representation by its real frequency counterpart
\begin{align}
\label{eq:T-matrix}
T_2 (\omega, {\bf q}) &= - \frac{2 \pi}{m} \frac{1}{\ln a_{\rm 2} \sqrt{- m (\omega - \varepsilon_{\bf q}/2) - i 0}} .
\end{align}
The remaining momentum integration in Eq.~\eqref{eq:self-energy-expansion} is performed numerically. We emphasize that the analytic continuation is performed analytically and the numerical calculation determines the self-energy at real frequency. The imaginary part of Eq.~\eqref{eq:self-energy-expansion} is computed directly using {\tt Mathematica}. The real part is then obtained by a numerical Kramers-Kronig transformation.

\section{Results for the spectral function}\label{sec:results}

\begin{figure}
\scalebox{0.7}{\includegraphics{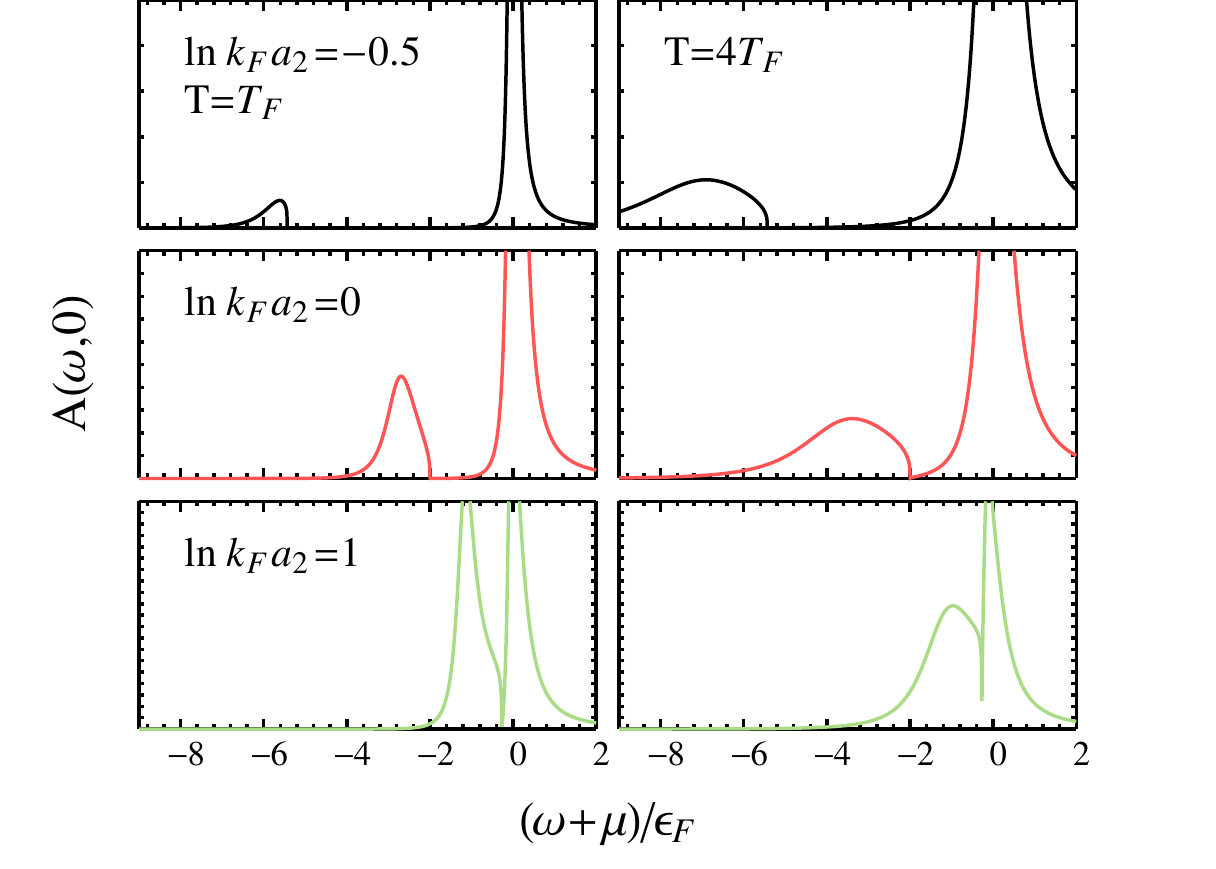}}
\caption{(Color online) Spectral function $A(\omega, 0)$ at fixed momentum $q=0$ and scattering lengths (rows) $\ln k_F a_{\rm 2} = -0.5,0$, and $1$. The first column is at $T=T_F$, the second at $T=4 T_F$.}
\label{fig:spectralslice}
\end{figure}

In this section, we present results for the spectral function calculated to leading order in the virial expansion. Figure~\ref{subfig:spectralfunction} shows the spectral function for a balanced Fermi gas at $T=T_F$ and $\ln k_F a_{\rm 2} = 0$. The spectral function exhibits a double-peak structure with two clearly distinguishable branches. The upper one, which we shall refer to as the quasiparticle branch, starts around zero frequency. The lower branch is associated with the existence of a bound state and starts at a threshold frequency $\omega_{\rm th} + \mu = - E_b$. The weight of both branches is shifted upwards with increasing momentum and displays a quadratic momentum dependence. The bound-state branch is strongly asymmetric: it quickly reaches its maximum below $\omega_{\rm th}$, but falls off slowly with decreasing frequency. This behavior is illustrated in Fig.~\ref{fig:spectralslice}, which shows the spectral function at fixed momentum ${\bf q} = 0$ as a function of temperature and scattering 
length. We see that the bound-state branch and the quasiparticle branch begin to merge as the scattering length is increased. With increasing temperature, the quasiparticle peak gets sharper and the effects of pairing become less relevant, a statement that will be made more precise in the following sections.

The asymmetric line shape of the lower branch drastically changes the form of the {\it occupied} part of the spectral function, which is the one measured experimentally. It is shown in Fig.~\ref{subfig:occspectralfunction}. The maximum of the lower branch increases quadratically at small momentum, but turns downwards at higher momentum. For small temperatures, the branch reaches its maximum at a momentum as low as the Fermi momentum. We emphasize that this is an effect of the thermal occupation of states and cannot be taken as a sign of a pseudogap. It is rather a generic feature of the occupied spectral function that is intrinsically linked to the enhanced short-range correlations in the system~\cite{schneider10}. The spectral weight at large momentum is the origin of high-momentum and high-frequency tails in the momentum distribution and the rf transition, respectively, two quantities that can be readily obtained from the spectral function. We examine the aforementioned properties closely in Sec.~\ref{sec:rfandmomdis}.
We conclude by studying the properties of the quasiparticle branch in Sec.~\ref{sec:qp}: we determine the quasiparticle dispersion as well as its effective mass and lifetime and compare them with exact results.

\begin{figure*}
\scalebox{0.7}{\includegraphics{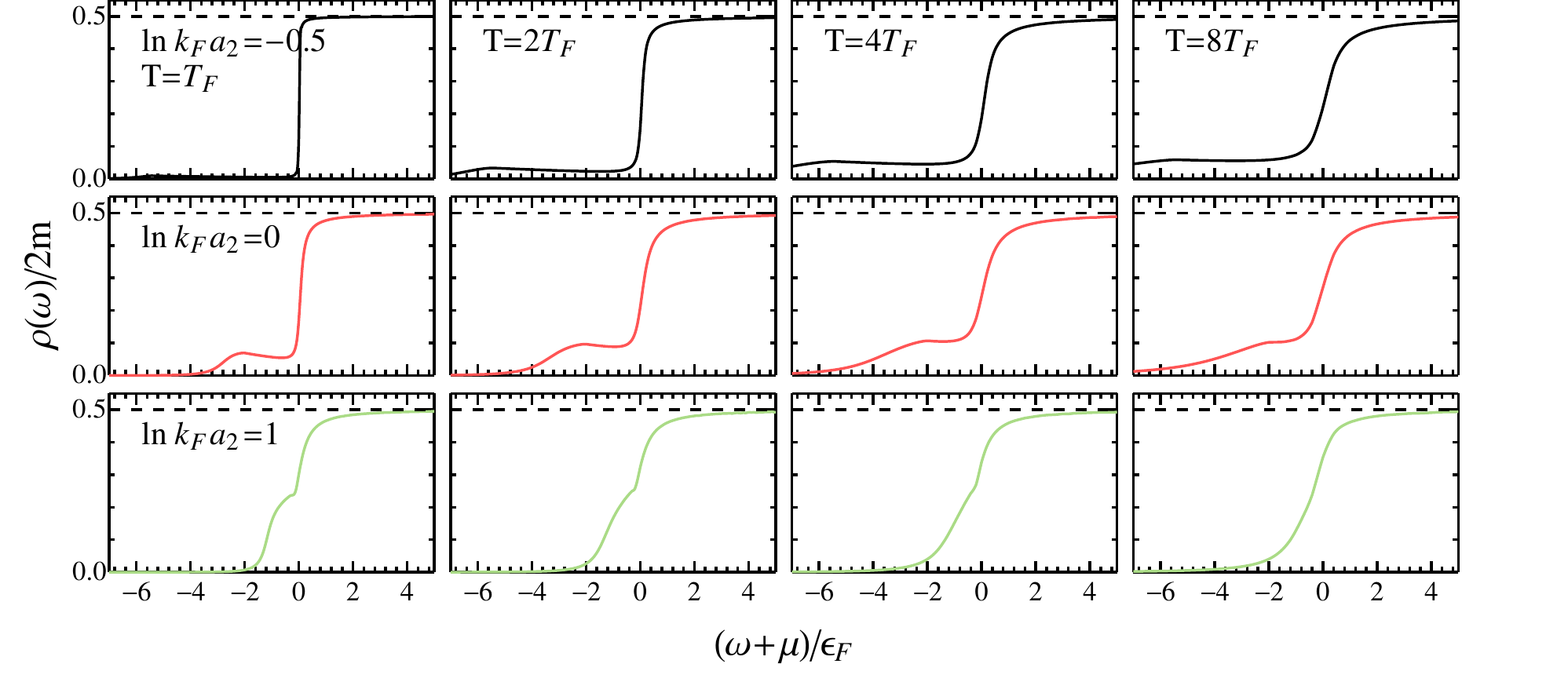}}
\caption{(Color online) Densities of states at temperatures (columns) $T/T_F=1,2,4$, and $8$ for scattering lengths (rows) $\ln k_F a_{\rm 2} = -0.5, 0,$ and $1.0$. The pairing gap is more pronounced as the scattering length decreases.}
\label{fig:dos}
\end{figure*}

\subsection{Density of states}\label{sec:dos}

Pairing effects are apparent in the density of states. Figure~\ref{fig:dos} shows the density of states for different scattering lengths and temperatures. At low temperature, a depletion of spectral weight around $\omega = -\mu$ is clearly visible, indicating that this effect is associated with the formation of a molecular bound state. This effect increases as the scattering length is lowered towards the BEC side of the crossover. For positive frequency, the density of states is very close to that of a free Fermi gas, which in 2D is simply given by
\begin{align}
\rho(\omega) &= m \, \Theta(\omega + \mu) .
\end{align}

For comparison, we also show the temperature evolution of the density of states in Fig.~\ref{fig:dos}. The dip in the spectral weight increases with decreasing temperature and resembles the density of states in the superfluid BEC regime shown in Fig.~\ref{fig:dossketch}. The transition occurs at a temperature scale $T^*$ that is in good agreement with our estimate~\eqref{eq:saha}; see Fig.~\ref{subfig:fugacity}. The two-body calculation does not show, however, the typical depletion for a pseudogap around $\omega=\mu$.

\subsection{Negative-frequency weight and universal relations}\label{sec:rfandmomdis}

\begin{figure*}
\subfigure[]{\scalebox{0.7}{\includegraphics{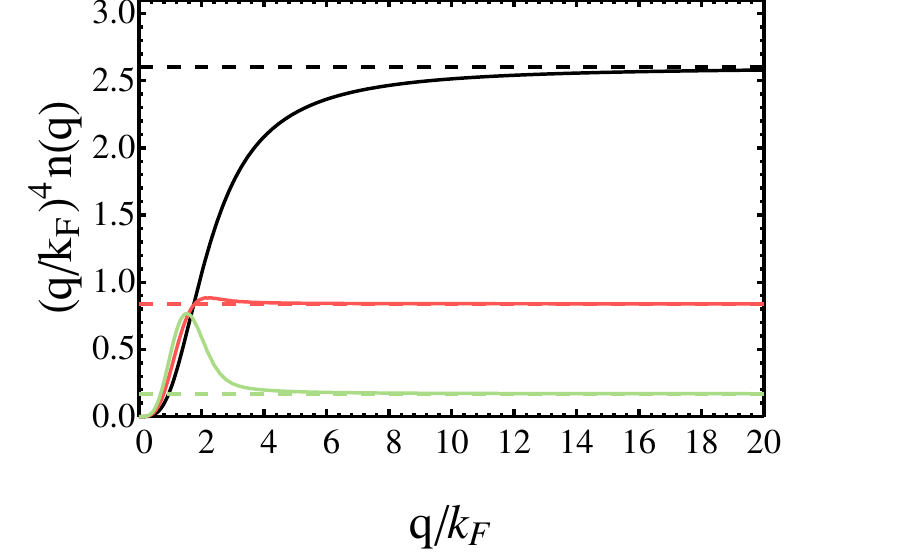}\label{fig:momdis}}}\hspace{-0.8cm}
\subfigure[]{\scalebox{0.7}{\includegraphics{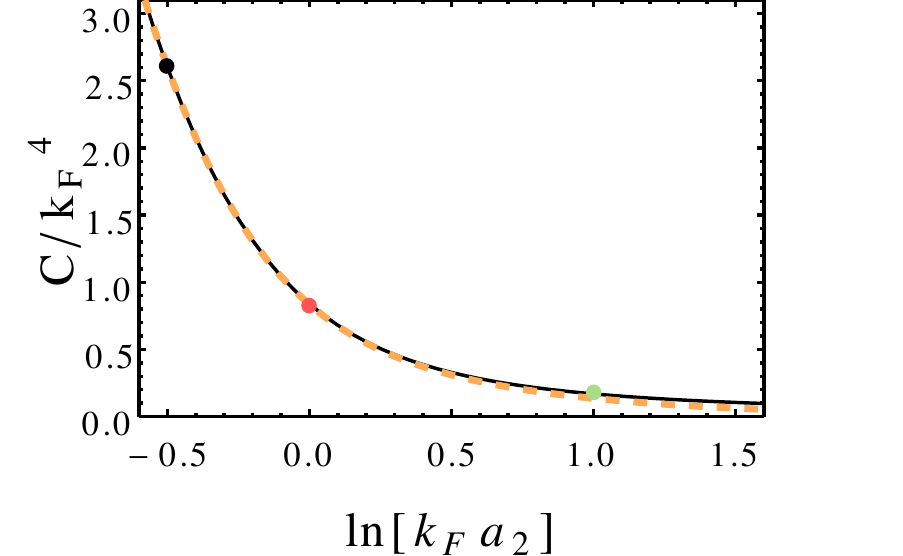}\label{fig:contact}}}\hspace{-0.8cm}
\subfigure[]{\scalebox{0.7}{\includegraphics{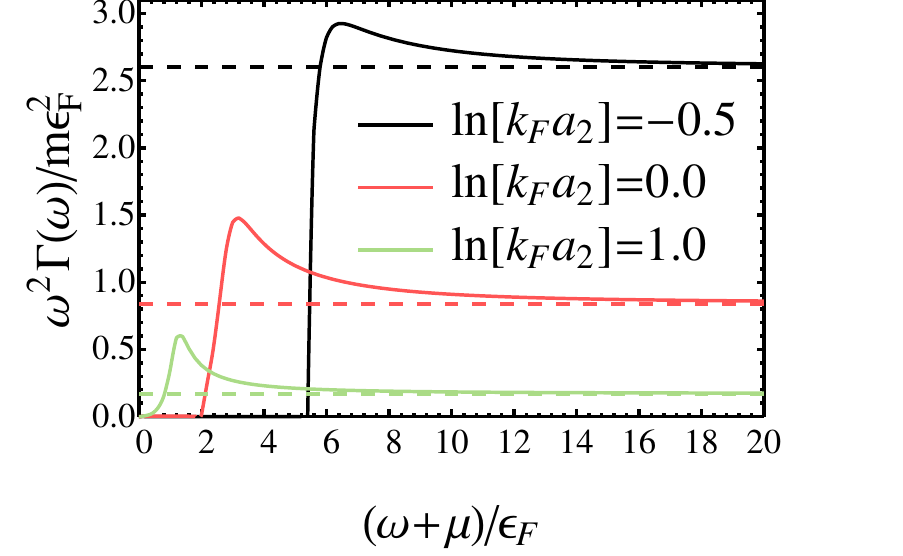}\label{fig:rfcontact}}}
\caption{(Color online) (a) Momentum distribution for $T=T_F$ and (top to bottom) $\ln k_F a_2 = -0.5$ (black), $\ln k_F a_2 =0$ (red, gray) and $\ln k_F a_2 =1$ (green, light gray) as obtained from the spectral function. To make the high-momentum tail visible, we multiplied the momentum distribution by $q^4$. Dashed lines are the values of the contact density as calculated from \eqref{eq:contact-from-b2}. (b) Contact density to leading order in the virial expansion at $T=T_F$. Points mark the values of the contact for the parameters used in figures (a) and (c). The dashed orange (gray) line is the contact for a gas of dimers. (c) rf spectra at the same parameter values. The spectra are multiplied by $\omega^2$ to extract the high-frequency tail. Dashed lines and parameters as in (a).}
\end{figure*}

We noted at the beginning of this section that the back-bending of the lower branch in the occupied spectral function is by itself not a sufficient sign of the pseudogap. It is rather a universal property of fermions with short-range interactions that exists independently of the phase or indeed temperature, and, in particular, it holds for any $N$-particle ensemble. At large momentum, the negative-energy weight gives the dominant contribution to the momentum distribution
\begin{align}
n_\sigma(q) &= \int \frac{d\omega}{2\pi} \, A_-(\omega, {\bf q}) , \label{eq:momdis}
\end{align}
resulting in a high-momentum tail $n_\sigma(q) \rightarrow {\cal C}/q^4$, where ${\cal C}$ is the so-called contact density~\footnote{We have defined our momentum distribution with an intensive normalization $\int \dfrac{d^2q}{(2\pi)^2} n_{\sigma} ({\bf q}) = n/2$, thus the contact density, not the (extensive) contact, is the coefficient of the high-momentum tail.}, which is a measure for the number of fermion pairs with opposite spins at short distances~\cite{tan08b,braaten08}. The contact density is related to the derivative of the grand canonical potential through the adiabatic relation~\cite{tan08b, werner12}
\begin{align}
 {\cal C} & = \left. 2 \pi m a_2 \frac{\partial \Omega / V}{\partial a_2} \right|_{T,\mu}
    = - 2 z^2 m^2 T^2 a_2 \left. \frac{\partial b_2}{\partial a_2} \right|_T + \mathit{O}(z^3) .\label{eq:contact-from-b2}
\end{align}
To leading order in $z$, the contact density can be determined from the second virial coefficient $b_2$ given in Eq.~\eqref{eq:beth-uhl}. The universal high-momentum tail for the momentum distribution is indeed obeyed by the virial expansion: in Fig.~\ref{fig:momdis}, the asymptotic behavior of the momentum distribution for different coupling strengths is shown. The high-momentum tail is clearly visible and fits well with the contact determined from the adiabatic theorem~\eqref{eq:contact-from-b2}. Quite generally, the scale at which the relation for the high-momentum tail holds is set by $q \gg \text{max}(1/\lambda_T,1/a_2,1/k_F)$, which explains that the green curve for $\text{ln} k_F a_2 = -0.5$ in Fig. \ref{fig:momdis} saturates much later than for the two larger values of $a_2$. In Fig.~\ref{fig:contact},
 we report the contact density to second order as obtained from the adiabatic relation~\eqref{eq:contact-from-b2}. The dashed line denotes the bound state contribution
\begin{align}
\mathcal{C}^{\rm bound} &= \frac{4 \pi}{a_{\rm 2}^2} \, n_d, \label{eq:contactboundstate}
\end{align}
which dominates for most scattering lengths. 
The bound state contribution is a homogeneous function of the fermion density: ${\cal C}^{\rm bound}  \sim n^2$. Although small compared to the bound state part, the remaining interaction contribution violates this simple scaling behavior. This affects the oscillation frequency of collective modes at low temperature~\cite{hofmann12,taylor12,chafin13}.

\begin{figure}[b]
\scalebox{0.65}{\includegraphics{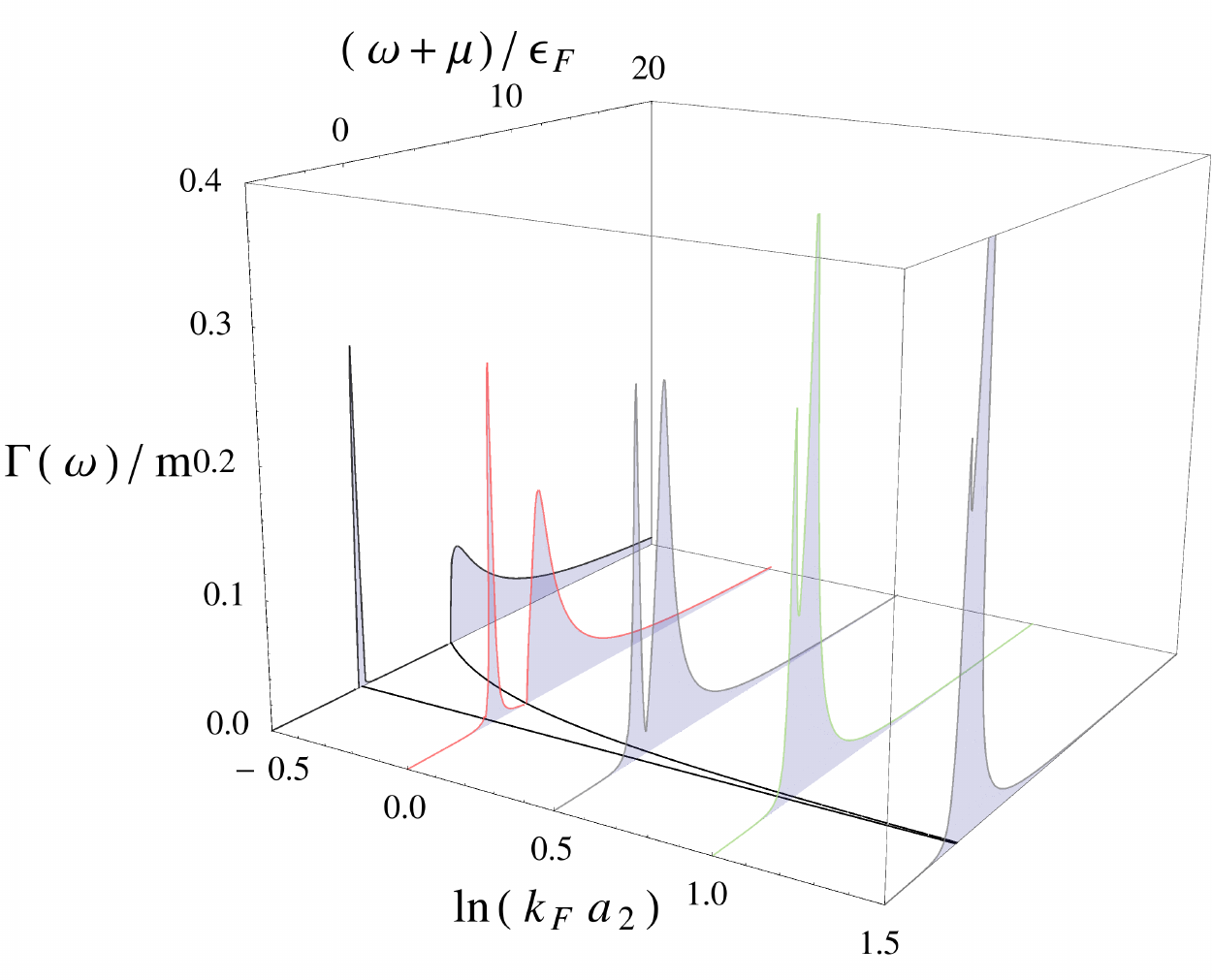}}
\caption{(Color online) rf spectrum at $T=T_F$ for scattering lengths  $\ln k_F a_2 = -0.5$, $0.0$, $0.5, 1.0$, and $1.5$. The peak at zero frequency is the transition from the quasiparticle branch. For energies larger than the bound-state energy (indicated by the black curve), there is an extended spectral weight due to bound-free transitions, which merges with the zero-frequency peak at large scattering length on the BCS side of the crossover.}
\label{fig:rf}
\end{figure}

The spectral function also determines the total rf transition rate. Provided that the final state does not interact with the two other species and is initially not populated, the transition rate is given by~\cite{haussmann09}
\begin{align}
\label{eq:rfdef}
\Gamma(\omega) &= \Omega^2 \int \frac{d^2q}{(2\pi)^2} \, A_-(\varepsilon_{\bf q} - \omega - \mu, {\bf q}),
\end{align}
which is just an integral over the occupied part of the spectral function evaluated at the free particle energies shifted by the transition frequency. In the following, we set the Rabi frequency $\Omega$ of the transition equal to $1$, resulting in the normalization
\begin{align}
 \int \frac{d\omega}{2\pi} \, \Gamma(\omega) = \frac{n}{2} .
\end{align}
At large frequency, the rf transition rate displays a universal tail~\cite{langmack12,sommer12}
\begin{align}
\Gamma(\omega) &\to \frac{{\cal C}}{4 m  \omega^2}.
\end{align}
It should be noted that final-state interactions introduce a logarithmic scaling violation $\sim 1/\omega^2 \ln^2\omega$~\cite{langmack12}. From Eq.~\eqref{eq:rfdef}, we see that the high-frequency tail is a direct consequence of the incoherent negative weight at large momentum, just as for the momentum distribution. The asymptotic form is again very well reproduced by the virial expansion as can be seen from Fig.~\ref{fig:rfcontact} which shows the asymptotic behavior of the transition rate at $T=T_F$. In Fig.~\ref{fig:rf}, we report the corresponding rf spectra. The peak at $\omega=0$ corresponds to transitions from the quasiparticle branch. The large incoherent weight starting at the binding energy $E_b$ corresponds to excitations that break up a dimer. For smaller binding energies (larger $\ln k_F a_2$), the bump of the dimer-free transition becomes sharper and begins to overlap with the free-free peak. This is to be expected, as for $E_b\rightarrow 0$, the spectrum needs to reproduce the one of free 
particles, which corresponds to a peak at $\omega=0$. Due to thermal excitations, the peak has always a finite width.

It is indeed no coincidence that the universal relations are obeyed by the cluster expansion. As stressed at the beginning of this section, the exact relations hold for any contact interacting system of $N$ particles. A simple power counting in the fugacity shows that each order in the virial expansion will reproduce the asymptotic tail at the same order of the contact. The fact that our calculation reproduces the universal relations with high accuracy is not only a stringent test of our computation but also shows that the quantum cluster expansion captures the correct short-time and distance structure of the system.

\begin{figure*}
\subfigure[]{\scalebox{0.6}{\includegraphics{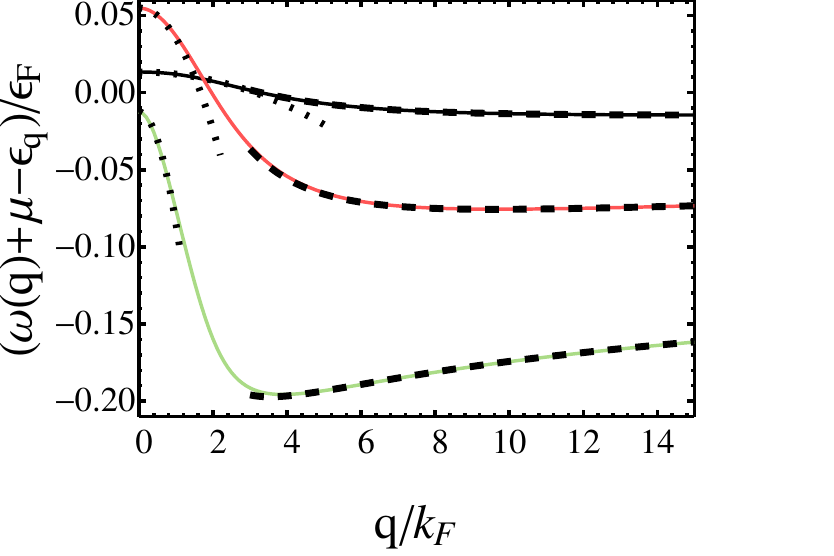}}\label{fig:qpenergy}}\hspace{-0.9cm}
\subfigure[]{\scalebox{0.6}{\includegraphics{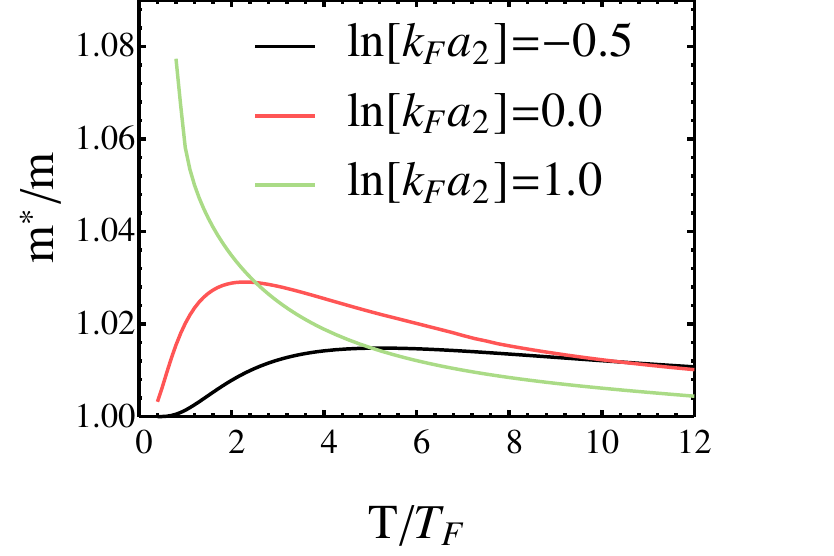}}\label{fig:qpeffmass}}\hspace{-0.9cm}
\subfigure[]{\scalebox{0.6}{\includegraphics{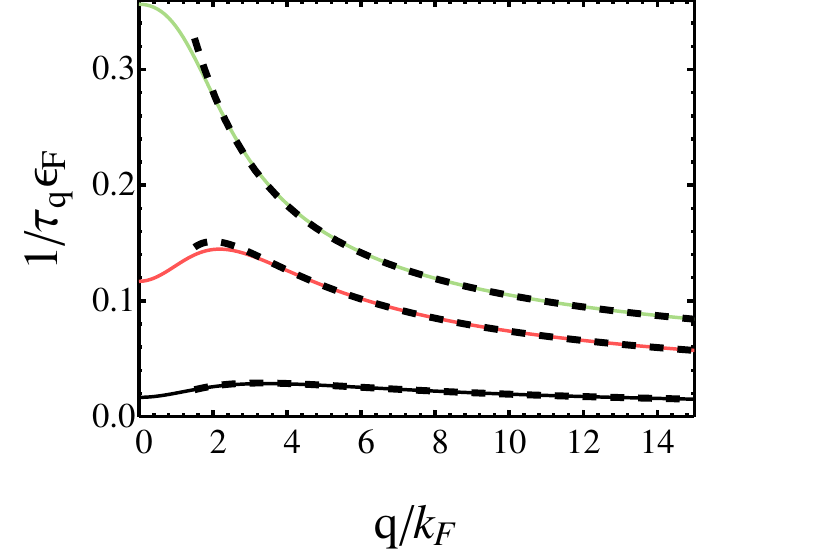}}\label{fig:qplifetime}}\hspace{-0.9cm}
\subfigure[]{\scalebox{0.6}{\includegraphics{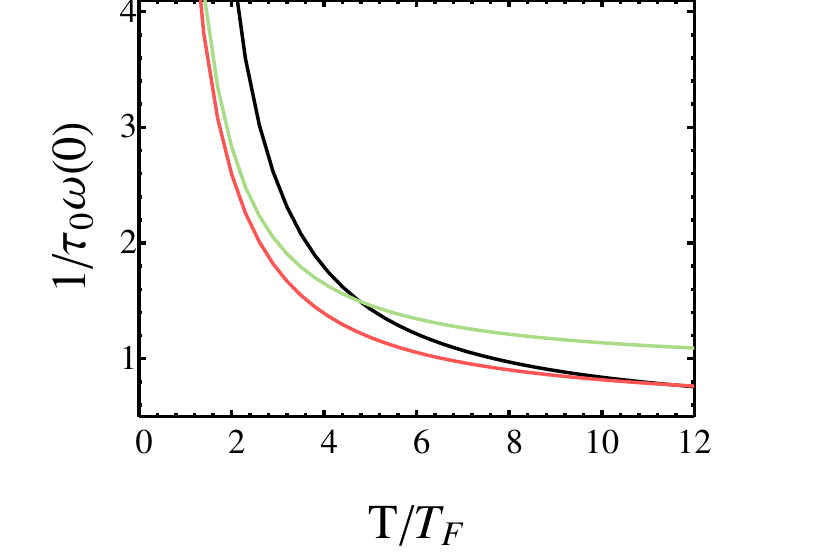}}\label{fig:qpdecay}}
\caption{(Color online) (a) Quasiparticle dispersion relation, (b) effective mass, and (c) inverse quasiparticle lifetime. The parameters are $T=T_F$ and $\ln k_F a_{\rm 2} = -0.5$ (black), $0.0$ (red, gray) and $1$ (green, light gray). The dashed lines indicate the asymptotic forms~\eqref{eq:asyenergy} and~\eqref{eq:asylifetime}. The dotted lines are the low-momentum limits of the dispersion relation with an effective mass~\eqref{eq:effmass}. (d) Ratio of inverse quasiparticle lifetime and energy $1/\tau_0 \omega(0)$ at zero momentum.}
\label{fig:qp}
\end{figure*}

\subsection{Quasiparticle branch}\label{sec:qp}

Let us now consider the quasiparticle branch. The spectral weight of this branch is centered around the quasiparticle energy $\omega({\bf q})$, which is given by the pole in the Green's function~\cite{landau10}
\begin{align}
\omega({\bf q}) - \varepsilon_{\bf q} + \mu - {\rm Re} \, \Sigma(\omega({\bf q}), {\bf q}) &= 0 .
\end{align}
In Fig.~\ref{fig:qpenergy}, we plot the dispersion relation $\omega({\bf q}) + \mu$ as a function of the momentum $q/k_F$ for $T=T_F$ and three different scattering lengths $\ln k_F a_{\rm 2} = -0.5,0,$ and $1$.  The dispersion relation starts at
\begin{align}
\omega({\bf 0}) + \mu &= {\rm Re} \left. \Sigma(\omega,0) \right|_{\omega=\omega({\bf 0})} 
\end{align}
and is quadratic at small momentum with an effective mass
\begin{align}
\frac{m^*}{m} &= \left. \frac{1 - \dfrac{\partial {\rm Re} \, \Sigma}{\partial \omega}}{1 + \dfrac{\partial {\rm Re} \, \Sigma}{\partial \varepsilon_{\bf q}}}  \right|_{\omega=\omega({\bf q}),q=0} . \label{eq:effmass}
\end{align}
Figure~\ref{fig:qpeffmass} shows the effective mass as a function of temperature for various scattering lengths. 
Our calculation indicates a slightly enhanced effective mass $m^*/m = 1.05$. At high temperature, the effective mass approaches the mass of the free Fermi gas. A small effective mass is quite typical even for strongly interacting Fermi gases at low temperature~\cite{combescot07}. In the zero-temperature limit, effective mass corrections in 2D are expected to get as large as $m^*/m \approx 1.5$ at $\ln k_Fa_{\rm 2}=0$ for the so-called attractive polaron~\cite{schmidt12}.

In the vicinity of $\omega({\bf q})$, the spectral function assumes a Lorentzian shape:
\begin{align}
\label{eq:lorentz}
A(\omega, {\bf q}) &= \frac{2/\tau_{\bf q}}{(\omega - \varepsilon_{\bf q} + \mu - {\rm Re} \, \Sigma(\omega,{\bf q}))^2 +(1/\tau_{\bf q})^2} .
\end{align}
The width of the Lorentzian, which is determined by the imaginary part of the self-energy, describes the rate at which a momentum state scatters into other momentum states~\cite{giuliani05}
\begin{align}
\frac{1}{\tau_{\bf q}} &= - {\rm Im} \, \Sigma(\omega({\bf q}), {\bf q}) .
\end{align}
The inverse lifetime is plotted in Fig.~\ref{fig:qplifetime} at $T=T_F$ for various scattering lengths. The quasiparticles are well defined if the inverse lifetime is much smaller compared to the excitation energy:
\begin{align}
\frac{1}{\tau_{\bf q}} \ll \omega({\bf q}) .
\end{align}
In Fig.~\ref{fig:qpdecay}, we plot the ratio $1/\tau_{\bf 0} \omega({\bf 0})$. It vanishes with the logarithm of temperature at high temperature. This shows that in the limit of high temperature, the low-energy excitations of the two-dimensional Fermi gas are indeed well-defined quasiparticles, allowing for a kinetic description of its non-equilibrium properties. This is consistent with an analogous result for the three-dimensional unitary Fermi gas~\cite{enss11}.

As pointed out by Nishida~\cite{nishida12} for the three-dimensional Fermi gas at a large scattering length, the self-energy at large momentum and frequency is universal, i.e., it is independent of the microscopic details of the system's state. The functional form can be calculated analytically by means of an operator product expansion~\cite{braaten08,son10,hofmann11,nishida12}. The magnitude of this high-momentum and high-frequency tail is set by the density:
\begin{align}
\Sigma(\omega, {\bf q}) &= W_n(\omega, {\bf q}) n + \cdots , \label{eq:asymsigma}
\end{align}
where $W_n(\omega, {\bf q})$ is the so-called Wilson coefficient of the density. It is given by the two-particle scattering amplitude:
\begin{align}
W_n(\omega, {\bf q}) = T_2(\omega, {\bf q}) .
\end{align}
The relation~\eqref{eq:asymsigma} dictates the asymptotic form of the dispersion relation
\begin{align}
\omega({\bf q}) + \mu &= \varepsilon_{\bf q} + \frac{2 \pi n}{m} \, \frac{\ln \varepsilon_{\bf q} / 2E_b}{\ln^2 \varepsilon_{\bf q}/2E_b + \pi^2}   + \textit{O}\left(\frac{1}{q\ln^2 q}\right) \label{eq:asyenergy}
\end{align}
and the lifetime
\begin{align}
\frac{1}{\tau_{\bf q}} &=  \frac{2 \pi n}{m} \, \frac{\pi}{\ln^2 \varepsilon_{\bf q} /2 E_b + \pi^2}  + \textit{O}\left(\frac{1}{q\ln^3 q}\right) . \label{eq:asylifetime}
\end{align}
This relation is obeyed by the virial expansion and indicated by dashed lines in Figs.~\ref{fig:qpenergy} and \ref{fig:qplifetime}. In particular, since in the high-momentum limit
\begin{align}
\tau_{\bf q} \, \omega({\bf q}) &= \frac{1}{\pi} \ln \frac{\varepsilon_{\bf q}}{2 E_b} \gg 1 ,
\end{align}
excitations at large momentum are always well-defined quasiparticles. This result holds at all temperatures.

In addition, Eq.~\eqref{eq:asymsigma} implies that the spectral function decays with the inverse power of frequency at high $\omega \gg q^2/2m$:
\begin{align}
A(\omega, {\bf q}) &= \frac{2\pi n}{m}\, \frac{1}{\omega^2} \frac{2\pi}{\ln^2 \omega/E_b + \pi^2}  + {\cal O}(1/\omega^3) .
\end{align}
To the best of our knowledge, this is a novel universal relation for fermions with short-range interactions in 2D. The next-to-leading order is proportional to the contact parameter. Calculating this contribution would require the inclusion of three-particle processes.

\section{CONCLUSION AND OUTLOOK}\label{sec:conclusion}

In conclusion, we have calculated the spectral function of a spin-balanced two-dimensional Fermi gas with short-range interactions to leading order in a virial expansion. This order takes into account two-particle effects and reproduces the salient features of the spectral function, which is dominated by a quasiparticle branch and a branch at lower energy that is associated with the two-particle bound state. Our results give a good qualitative description of recent experiments~\cite{feld11}.

It turns out that the virial expansion can be applied to temperatures as low as the Fermi temperature $T_F$, a regime where pairing affects the single-particle spectrum and the density of states. While the onset of a pairing gap is visible in the density of states, it is interesting to note that the back-bending of the lower branch of the occupied spectral function with increasing momentum does not appear to be a sufficient sign of a pseudogap, for this is not seen in the full spectral function. It is an artifact of combining the finite width of the lower branch with the thermal occupation that weighs the measured spectra to lower frequencies.

The spectral function is related to various observable quantities, notably the momentum distribution and the rf transition rate, both of which were calculated in this paper, and excellent agreement with exact universal results was found. Furthermore, we analyzed the quasiparticle branch and determined the quasiparticle properties. Effective-mass corrections are found to be very small, while the lifetime of the quasiparticle branch approaches very large values as the temperature is increased.

The present work could be straightforwardly extended in several ways, for example by including the effects of harmonic confinement in two- or quasi-two-dimensional geometries. It would also be interesting to extend the range of validity to even lower temperatures by performing the quantum cluster expansion to next-to-leading order, which takes into account three-particle processes.

\textit{Note added:} Recently, we became aware of Ref.~\cite{wave13}, where some of our results have been derived independently.

\begin{figure}[t]
\subfigure[\label{subfig:virial-prop}]{\raisebox{0.25cm}{\scalebox{0.4}{\includegraphics{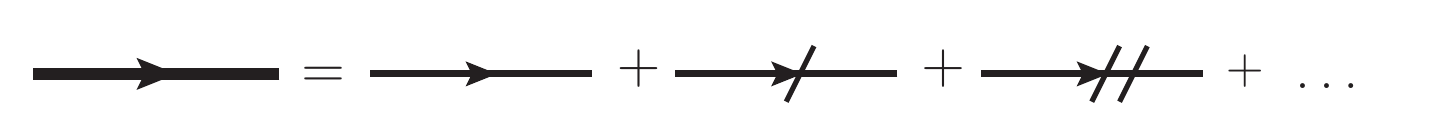}}}}
\subfigure[\label{fig:self-energy-diagram}]{\scalebox{0.4}{\includegraphics{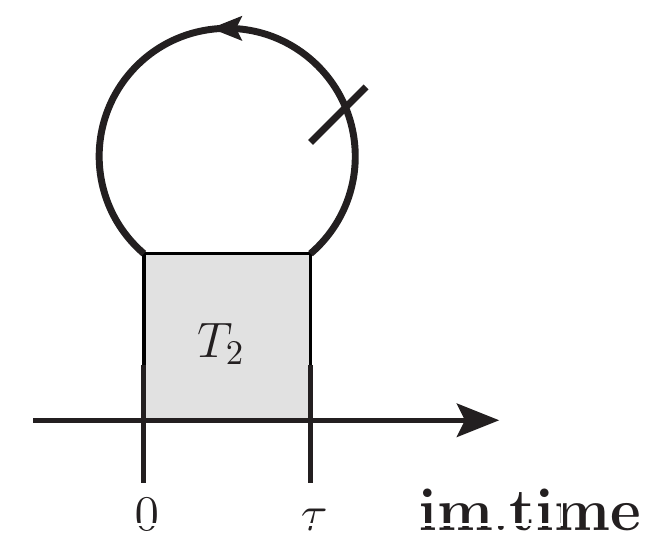}}}
\caption{(a) Diagrammatic representation of Eq.~\eqref{eq:virial-prop-series}. The bare propagator (thick line) is a series of expanded propagators $G^{(n)}_0$ (continuous thin lines). The number of slashes counts the expansion order $n$ in the fugacity. (b) Diagram for the self-energy expanded to first order in the fugacity $z$.}
\label{fig:td-virial-diagrams}
\end{figure}

\begin{acknowledgments}
M.B. is supported by the DFG research unit ``Strong Correlations in Multiflavor Ultracold Quantum Gases.'' We thank Marianne Bauer, Richard Schmidt, and Wilhelm Zwerger for useful discussions.
\end{acknowledgments}

\appendix

\section{Diagrammatic Formalism} \label{app:diagrams}

In this appendix, we derive the leading-order contribution~\eqref{eq:self-energy-expansion} to the virial expansion of the self-energy using a diagrammatic formalism. In Ref.~\cite{leyronas11}, the relation \eqref{eq:density-grandcanonical} for the number density was taken as a starting point to calculate the virial coefficients of a three-dimensional contact-interacting Fermi gas up to third order using a diagrammatic approach. Some resummation schemes, such as the $T$-matrix approximation, seem to reproduce the results of the leadingorder virial expansion at high temperature~\cite{combescot06}. It should be noted that the diagrammatic formalism is not restricted to contact interactions, but can also be applied to other systems such as the electron gas~\cite{hofmann13}. Here, we apply the same formalism to the two-dimensional Fermi gas. The starting point is the free-fermion propagator, which in imaginary time is given by
\begin{align}
G_0 ( \tau , {\bf q})&= e^{-(\varepsilon_{\bf q} - \mu) \tau}\left( f (\varepsilon_{\bf q} - \mu) - \Theta (\tau) \right),
\end{align}
where $\Theta (\tau)$ denotes the Heaviside function. Expanding the Fermi distribution $f(\varepsilon_{\bf q} - \mu)$
with respect to the fugacity $z$ in the above equation yields
\begin{align}
G_0 ( \tau , {\bf q}) & = e^{\mu \tau} \sum_{n \geq 0} G^{(n)}_0 (\tau, {\bf q})\,z^n , \label{eq:virial-prop-series}
\end{align}
where
\begin{align}
G^{(n)}_0 (\tau, {\bf q}) & = \begin{cases} -\Theta (\tau) e^{-\varepsilon_{\mathbf{q}} \tau}              & n=0
                                \\ (-1)^{n-1} e^{-\varepsilon_{\bf q} \tau} e^{-n \beta \varepsilon_{\bf q}} & n\geq1
                                \end{cases} .
\end{align}
Following Leyronas~\cite{leyronas11}, we depict the $n$-th order term $G^{(n)}_0$ diagrammatically by a line that is slashed $n$ times [see Fig.~\ref{subfig:virial-prop}].

A given Feynman diagram with $G^{(n)}_0$ appearing $N_n$ times is of order $\sum_{n} n N_n$ in the fugacity. Since $G^{(0)}_0$ is a retarded Green's function, it is not allowed to propagate backwards in imaginary time. The leading order in $z$ is thus given by the diagram with the least number of advanced propagators. The self-energy to first order in the fugacity can be inferred directly from Fig.~\ref{fig:self-energy-diagram}, which is the only one-particle irreducible diagram containing only one backward-propagating propagator. It describes the interaction with a single particle-hole pair:
\begin{align}
 &\Sigma^{(1)}(i \omega_n, {\bf q}) \nonumber \\
 &= z \int_0^{\beta} d\tau \int \frac{d^2k}{(2\pi)^2} e^{i \omega_n \tau} e^{\mu \tau} e^{-\varepsilon_{\bf q} (\beta - \tau)} T_2( \tau, {\bf k + q}) ,
\end{align}
which gives the result in Eq.~\eqref{eq:self-energy-expansion}. The $T$-matrix $T_2$ is the ladder series of all forward-propagating lines. It is of zeroth order in $z$ and equivalent to the vacuum $T$ matrix.
\bibliography{bib}

\end{document}